\documentclass[altaffilletter,aps,nofootinbib,twocolumn,prd,eqsecnum,preprintnumbers,superscriptaddress,10pt,floatfix]{revtex4-1}
\pdfoutput=1
\usepackage[caption=false]{subfig}
\usepackage{graphicx}
\usepackage{amsmath}
\usepackage{amsfonts}
\usepackage{mathtools}
\usepackage{amssymb}
\usepackage{enumerate}
\usepackage{xcolor}
\usepackage{bm}
\usepackage{mathrsfs}
\usepackage{epstopdf}
\usepackage{url}
\usepackage{footnote}
\usepackage{textcomp}
\usepackage{dsfont}
\usepackage{ulem}
\usepackage{hyperref}
\usepackage{enumerate}   
\usepackage{appendix}
\usepackage{textcomp}
\usepackage{tipa}

\makeatletter
\newcommand*{\rom}[1]{\expandafter\@slowromancap\romannumeral #1@}
\makeatother

\begin{document}

\title{A novel test of gravity via black hole eikonal correspondence}

\author{Che-Yu Chen}
\email{b97202056@gmail.com}
\affiliation{Institute of Physics, Academia Sinica, Taipei 11529, Taiwan}

\author{Yu-Jui Chen}
\email{r10222095@ntu.edu.tw}
\affiliation{Department of Physics and Center for Theoretical Physics, National Taiwan University, Taipei 10617, Taiwan}

\author{Meng-Yuan Ho}
\email{r11222007@ntu.edu.tw}
\affiliation{Department of Physics and Center for Theoretical Physics, National Taiwan University, Taipei 10617, Taiwan}

\author{Yung-Hsuan Tseng}
\email{anna424.tseng@gmail.com}
\affiliation{Department of Physics, National Tsing Hua University, Hsinchu 30013, Taiwan}

\begin{abstract}
When adopted in black hole spacetimes, geometric-optics approximations imply a mapping between the quasinormal mode (QNM) spectrum of black holes in the eikonal limit and black hole images. In particular, the real part and the imaginary part of eikonal QNM frequencies are associated with the apparent size and the detailed structure of the ring images, respectively. This correspondence could be violated when going beyond general relativity. We propose a novel method to test the eikonal correspondence via the comparison of two sets of observables from a nonrotating black hole, one extracted from QNM spectra and the other from the lensed photon rings on the image plane. Specifically, the photon ring observables robustly capture the information of the black hole spacetime itself regardless of the surrounding emission models. Therefore, the proposed test of eikonal correspondence can be validated in quite broad scenarios.   

\end{abstract}

\maketitle

\section{Introduction}

With the detection of gravitational waves emitted from binary black hole mergers \cite{LIGOScientific:2016aoc} and the observations of the images of supermassive black holes \cite{EventHorizonTelescope:2019dse,EventHorizonTelescope:2022wkp}, we have been ushered in a new era in which probing strong gravity regimes, such as the spacetime near black hole horizons, becomes feasible. These scientific advancements also provide us with the possibility of testing the validity of general relativity (GR) on very extreme scales. In particular, one could utilize these black hole observations to test the fundamental principles or symmetries dictated by GR, such as the black hole no-hair theorem \cite{Isi:2019aib,Johannsen:2010ru}, circularity \cite{Long:2020wqj,Eichhorn:2021iwq,Delaporte:2022acp}, reflection symmetry \cite{Chen:2020aix,Fransen:2022jtw,Chen:2022lct}, etc. 

One interesting property inherent in GR is the eikonal correspondence between the wave and ray quantities in generic curved spacetimes \cite{Ferrari:1984zz,Hod:2009td}. The correspondence can be roughly understood by noting that high-frequency waves propagate in a similar manner as photons. More precisely, the wave dynamics in the so-called geometric-optics approximations, which assume that the wavelength of the propagating waves is much shorter than any other length scales in the system, resembles that of light rays. When applied to black hole spacetimes, the eikonal correspondence could have implications on the aforementioned two state-of-the-art black hole observations. More explicitly, the black hole quasinormal mode (QNM) frequencies \cite{Kokkotas:1999bd,Berti:2009kk,Konoplya:2011qq}, which play important roles in the post-merger phase of a binary coalescence, can be mapped to the properties of photons propagating near the bound photon orbits around the black hole \cite{Cardoso:2008bp,Dolan:2010wr,Yang:2012he,Li:2021zct}. This correspondence is more justified for QNMs with higher oscillation frequencies, i.e., in the eikonal limit, and can be extended further to map to some features in the ring-like images cast by the black hole \cite{Stefanov:2010xz,Jusufi:2019ltj,Jusufi:2020dhz,Cuadros-Melgar:2020kqn,Yang:2021zqy}. Roughly speaking, the real and the imaginary parts of eikonal QNM frequencies correspond to the apparent ring size and the detailed ring structure on the image plane.

Assuming the fulfillment of the correspondence, its identification in different black hole spacetimes typically relies on the existence of some symmetries of the spacetime \cite{Cardoso:2008bp,Dolan:2010wr,Yang:2012he,Li:2021zct,Assumpcao:2018bka},  such as the separability of the angular and radial sectors in geodesic equations and wave equations. Recently, the correspondence for black hole spacetimes with less separability has been identified \cite{Chen:2022ynz}. Some natural questions then arise: What if the correspondence is violated, and can it be tested using black hole observations? In fact, the fulfillment of eikonal correspondence is associated with the assumption that matter fields are minimally coupled to gravity in GR. In the presence of nonminimal couplings between photons and other degrees of freedom, photons and those degrees of freedom may propagate in a nontrivial manner such that the eikonal correspondence is violated \cite{Chen:2021cts,Chen:2019dip}. However, an explicit example studying the charged black holes in nonlinear electrodynamics in Ref. \cite{Nomura:2021efi} illustrates that nonminimal coupling is not a necessary condition for the violation of eikonal correspondence, suggesting that the violation is triggered in a more general setting. In fact, the violation can also happen in models inspired by some putative quantum theories of gravity \cite{Konoplya:2017wot,Glampedakis:2019dqh,Silva:2019scu,Moura:2021eln,Bryant:2021xdh}. On the other hand, the presence of nonminimal coupling does not imply the violation of eikonal correspondence either \cite{Guo:2021enm}. In any case, the physics behind the violation of eikonal correspondence may be much more intricate than one naively expects. Therefore, the possibility of observationally testing eikonal correspondence deserves to be explored \cite{Li:2021mnx}, not only because of the potential future advancements of the observations related to both sides of the correspondence but also due to its theoretical relevance at a fundamental level. In short, any observational indication of the violation of eikonal correspondence can be a smoking gun of physics beyond GR.

In this paper, we propose a novel method of testing eikonal correspondence by observing the QNM spectrum and ring images of a black hole. We define a set of QNM observables as the ratio between the real and imaginary parts of QNM frequencies. On the other hand, we also define two sets of photon ring observables that, at this moment, still require significant improvements to resolve. Due to the strong lensing effects, photons emitted from a light source near a black hole could follow multiple paths and orbit the black hole different numbers of times before reaching the observer. These photons thus generate on the image plane a stack of discrete photon rings \cite{Gralla:2019xty,Gralla:2019drh}, whose resolution is beyond the reach of the current Event Horizon Telescope (EHT) collaboration, but may be partially resolved in the next-generation EHT (ngEHT) \cite{Blackburn:2019bly} or future space-based Very-Long-Baseline Interferometry (VLBI) missions \cite{Johnson:2019ljv,Haworth:2019urs,Gralla:2020nwp}. The ring observables can be extracted from these ring structures and, as we will show, can capture the features of the underlying black hole geometry. Most importantly, in the presence of eikonal correspondence and considering nonrotating black holes as toy models, the QNM and ring observables, in their certain limits, always converge to some critical exponent from opposite directions. This behavior of convergence seems to be insensitive to the emission models and the spacetime metric itself under consideration. Therefore, by directly comparing these observables, one can test the eikonal correspondence.


\section{Setup for the test}\label{sec.observable}

In this section, we will describe the basic setup of our black hole test. More explicitly, we will first introduce a critical exponent $\gamma$ which is solely defined by the properties of the spherical photon orbits around a black hole. This critical exponent $\gamma$ depends only on the black hole geometry, i.e., the metric of the black hole, not on the emission properties nearby. Then, we will introduce two sets of observables, one is defined via QNM frequencies and the other is defined through the structure of photon rings. These two sets of quantities are in principle potential observables in the future advance of gravitational wave detection and space-VLBI observations, respectively. The high-order photon ring observables naturally converge to the critical exponent $\gamma$. On the other hand, the QNM observables would converge to $\gamma$ only when the eikonal correspondence is satisfied. Although $\gamma$ itself is not an observable, the comparison between the two sets of observables still enables one to test the eikonal correspondence, as we will demonstrate later.

We consider a general static and spherically symmetric spacetime:
\begin{equation}
ds^2=-f(r)dt^2+\frac{g(r)}{f(r)}dr^2+r^2d\Omega_2^2\,,\label{metricsss}
\end{equation}
where the metric functions $f(r)$ and $g(r)$ are functions of the radial coordinate $r$. Also, $d\Omega_2^2$ is the line element on a 2-sphere. A generic black hole spacetime contains the following constant-$r$ surfaces: the event horizon $r_h$ that satisfies $f_h=0$, the photon sphere $r_{\textrm{ph}}$ that satisfies $2f_{\textrm{ph}}=r_{\textrm{ph}}f'_{\textrm{ph}}$, and the innermost stable circular orbit (ISCO) $r_{\textrm{ISCO}}$, which can be determined by the root of the following equation
\begin{equation}
\left[-3ff'/r+2(f')^2-ff''\right]_{\textrm{ISCO}}=0\,.
\end{equation}
The prime denotes the derivatives with respect to $r$. The subscript $x$ indicates that the quantities are evaluated at $r=r_x$. We remind that the photon sphere is a collection of spherical photon orbits around the black hole, which in the spherically symmetric case, has a single radius $r_{\textrm{ph}}$. The ISCO, on the other hand, defines the radius of circular timelike orbits on which the circular motion is marginally stable. The ISCO is typically regarded as the inner edge of accretion disks.

Having defined the radius of photon sphere $r_{\textrm{ph}}$, we can then define the Lyapunov exponent $\lambda$ and the angular velocity of photons on the photon sphere $\Omega_{\textrm{ph}}$ as follows \cite{Cardoso:2008bp}:
\begin{equation}
\lambda=\frac{1}{\sqrt{2}}\sqrt{-\frac{r_{\textrm{ph}}^2f_{\textrm{ph}}}{g_{\textrm{ph}}}\left(\frac{f}{r^2}\right)_{\textrm{ph}}''}\,,\qquad\Omega_{\textrm{ph}}=\frac{f^{1/2}_{\textrm{ph}}}{r_\textrm{ph}}\,.
\end{equation} 
The critical exponent is defined by their ratio:
\begin{equation}
    \gamma\equiv\frac{\lambda}{\Omega_\textrm{ph}}\,.\label{criticalgamma}
\end{equation}
For a Schwarzschild black hole whose metric functions are $g(r)=1$ and $f(r)=1-2M/r$ with its mass $M$, one can easily check that $\gamma=1$. 

\subsection{QNM observables}

The first observable we will consider in this test is defined by the black hole QNM frequencies. Being complex-valued, the QNM frequencies of black holes can be decomposed into a real part and an imaginary part as $\omega=\omega_R+i\omega_I$. For nonrotating black holes, the QNMs can be labeled by multipole numbers $l$ and overtones. Here, we will mainly focus on the fundamental modes since they have the longest decay time and are more astrophysically relevant. We then define our QNM observables as
\begin{equation}
    \gamma_l^{\textrm{QNM}}\equiv2l\frac{|\omega_I|}{\omega_R}\,,\label{QNMgamma}
\end{equation}
where their dependence on the multipole number $l$ is made explicitly in the lower index.

The calculations of black hole QNMs can be recast as a wave scattering problem around the black hole. For nonrotating black holes, the QNM spectrum can be determined by a master wave equation after suitable decomposition and field redefinition. Generically, the wave equations can be  written as
\begin{equation}
\frac{d^2\psi}{dr_*^2}+\left[\omega^2-U(r)\right]\psi=0\,,\label{waveeq}
\end{equation}
where $\psi$ stands for the fields scattered around the black hole and $r_*$ is the tortoise radius defined by $dr/dr_*=f(r)/g(r)^{1/2}$. The event horizon $r=r_h$ is mapped to $r_*\rightarrow-\infty$. The effective potential encodes all the information of the modes, and can be expressed as \cite{Glampedakis:2019dqh}
\begin{equation}
U(r)=f(r)\left[\frac{l(l+1)}{r^2}\alpha(r)-\frac{6M}{r^3}\zeta(r)\right]\,,
\end{equation}
where $\alpha(r)$ and $\zeta(r)$ are assumed to be independent of the multipole $l$, but remain undetermined. Note that the spin of the propagating fields is essentially encoded in the function $\zeta(r)$. In the rest of the discussions, it is fair to assume that the effective potential $U(r)$ takes the standard form of an asymptotically flat black hole, i.e., it has a single peak and approaches zero toward the boundaries $r_*\rightarrow\pm\infty$.

The QNM frequencies can be evaluated by solving the wave equation \eqref{waveeq} after imposing proper boundary conditions. Their analytic expressions are feasible within eikonal approximations{\footnote{For convenience, we do not consider the additional branches of large-$l$ modes that may appear when either side of the boundaries differs from those of a typical asymptotically flat black hole, as recently discussed in Ref.~\cite{Konoplya:2022gjp}.}}. Up to the subleading eikonal order, the frequency can be analytically written as \cite{Glampedakis:2019dqh}:
\begin{align}
\omega_R&=\left(l+\frac{1}{2}\right)\sqrt{\tilde{U}_m}+O\left(l^{-1}\right)\,,\label{eq2.8}\\
\omega_I&=-\frac{1}{2}\left(\frac{dr}{dr_*}\right)_m\sqrt{\frac{|\tilde{U}''_{m}|}{2\tilde{U}_m}}+O\left(l^{-1}\right)\,,\label{eq2.9}
\end{align}
where $\tilde{U}(r)=f(r)\alpha(r)/r^2$, and the subscript $m$ means that the quantities are evaluated at the peak of the potential $\tilde{U}(r)$, say, $r_m$. Therefore, in the eikonal limit, the QNM observables \eqref{QNMgamma} can be written as
\begin{equation}
\gamma_l^{\textrm{QNM}}\approx\left(\frac{2l}{2l+1}\right)\sqrt{-\frac{r_m^4}{2g_m\alpha_m^2}\left(\frac{f\alpha}{r^2}\right)''_{m}}\,.\label{gammaqnm210}
\end{equation}
If $\alpha(r)=1$, the peak of the potential $r_m$ is precisely at $r_{\textrm{ph}}$. In this case, we have
\begin{equation}
\gamma_l^{\textrm{QNM}}\approx\left(1-\frac{1}{2l}\right)\gamma\,,\label{gammaqnm211}
\end{equation}
implying that $\gamma_l^{\textrm{QNM}}\rightarrow\gamma$ when $l\rightarrow\infty$. Note that $\zeta(r)$ does not contribute to the right-hand side of Eq.~\eqref{gammaqnm210}. As a consequence, the function $\alpha(r)$ directly controls the eikonal correspondence. If $\alpha(r)$ is not a constant, $r_m$ would not be located at the photon sphere and $\gamma_l^{\textrm{QNM}}$ would not converge to $\gamma$ even in the eikonal limit.

One important observation from Eq.~\eqref{gammaqnm211} is that even when the eikonal correspondence holds, contributions of subleading eikonal orders make the QNM observables smaller than the critical exponent, i.e., $\gamma_l^\textrm{QNM}<\gamma$ for $l<\infty$. This observation will be supported later by our numerical calculations for different black hole models, and will play a crucial role in the validity of our test.

\subsection{Photon ring observables}

As mentioned in the Introduction, there are multiple photon trajectories connecting the observer on the earth to a source emission located near or behind the black hole. Photons emitted from a given light source but reaching the observer along different trajectories, which typically rotate around the black hole  different numbers of times, would generate a sequence of discrete photon rings \cite{Gralla:2019xty,Gralla:2019drh}. These photon rings are conveniently labeled by the winding number $n$, which stands for the number of half-orbits the photons rotate around the black hole before reaching the observer. As $n$ increases, the photon rings on the image plane would exponentially converge to the theoretical critical curve ($n=\infty$), which is the impact parameter of the spherical photon orbits around the black hole. The images contributed from direct emission, i.e., $n=0$ highly depend on the size of the emission region and its profile. However, in the advancement of the future space-VLBI and ngEHT observations, we may hopefully resolve higher-order rings, which would be less sensitive to the emission models and encode more information about the black hole geometry \cite{Wielgus:2021peu,Broderick:2021ohx,Kocherlakota:2022jnz}.

Consider an optically and geometrically thin disk model and assume that the black hole is being observed face-on. In this case, each photon ring appears as a circular ring shape with outer and inner radii determined by the impact parameters of photons emitted from the outer and inner edges of the disk, respectively. We denote the outer radius and the width of each ring as $b_n$ and $w_n$, respectively. Then, we can define two photon ring observables \cite{Kocherlakota:2022nhehttalk,Kocherlakota:2023qgo}
\begin{equation}
    \gamma_n^\textrm{w}\equiv\frac{1}{\pi}\ln{\frac{w_n}{w_{n+1}}}\,,\quad
    \gamma_n^\textrm{b}\equiv\frac{1}{\pi}\ln{\frac{b_n-b_{n+1}}{b_{n+1}-b_{n+2}}}\,.\label{ringwb}
\end{equation}
The first observable $\gamma_n^\textrm{w}$ is defined through the width ratio of two successive rings. The second observable $\gamma_n^\textrm{b}$ is a variant of the one proposed in \cite{Kocherlakota:2022nhehttalk,Kocherlakota:2023qgo} and is determined by the outer radii of three successive rings. Note that $\gamma_n^\textrm{b}$ does not require the resolution of any inner ring radius. As we will show numerically later and analytically in appendix~\ref{appendix1}, these two observables converge very quickly to $\gamma$ as $n$ increases. In fact, they are already very close to $\gamma$ at $n=1$ for different disk models with various sizes. In addition, our numerical calculations seem to indicate that these observables would approach from above to $\gamma$ when increasing $n$ for various black hole metrics{\footnote{This can be proven analytically when the black hole metric differs slightly from Schwarzschild one. See appendix~\ref{appendix1}.}}. This fact, combined with the fact that $\gamma_l^\textrm{QNM}<\gamma$, would enable us to test the eikonal correspondence by simply comparing the measurements of $\gamma_l^\textrm{QNM}$ and the ring observables.

\section{Examples}\label{sec.examples}
In this section, we consider some simple models to demonstrate how the two sets of observables defined in sec.~\ref{sec.observable} can be used to test eikonal correspondence. We first show how these observables change when the spacetime varies from the Schwarzschild one. To calculate the ring observables, two different disk models (disk modes 1 and 2) are considered. As Eq.~\eqref{ringwb} shows, the photon ring observables are defined through the width ratio of two successive rings and the outer radii of three successive rings. The disk size, which is relevant in defining the ring edges, in disk model 1 is assumed to range from the ISCO to $r=100M$. On the other hand, we assume that disk model 2 has its outer edge at $r=10M$ and extends inward to the event horizon. As for the QNM observables, we consider the QNMs of a massless scalar field. The frequencies are evaluated using the Wentzel-Kramers-Brillouin method up to the 6th order \cite{Schutz:1985km,Konoplya:2003ii,Konoplya:2019hlu}. Two specific non-Schwarzschild metrics will be considered as examples. After that, the section will end with an explicit illustration on how to test the eikonal correspondence using these observables.

\subsection{Reissner-Nordstr\"om spacetime}

\begin{figure*}[!ht]
    \centering
    \includegraphics[scale=0.549]{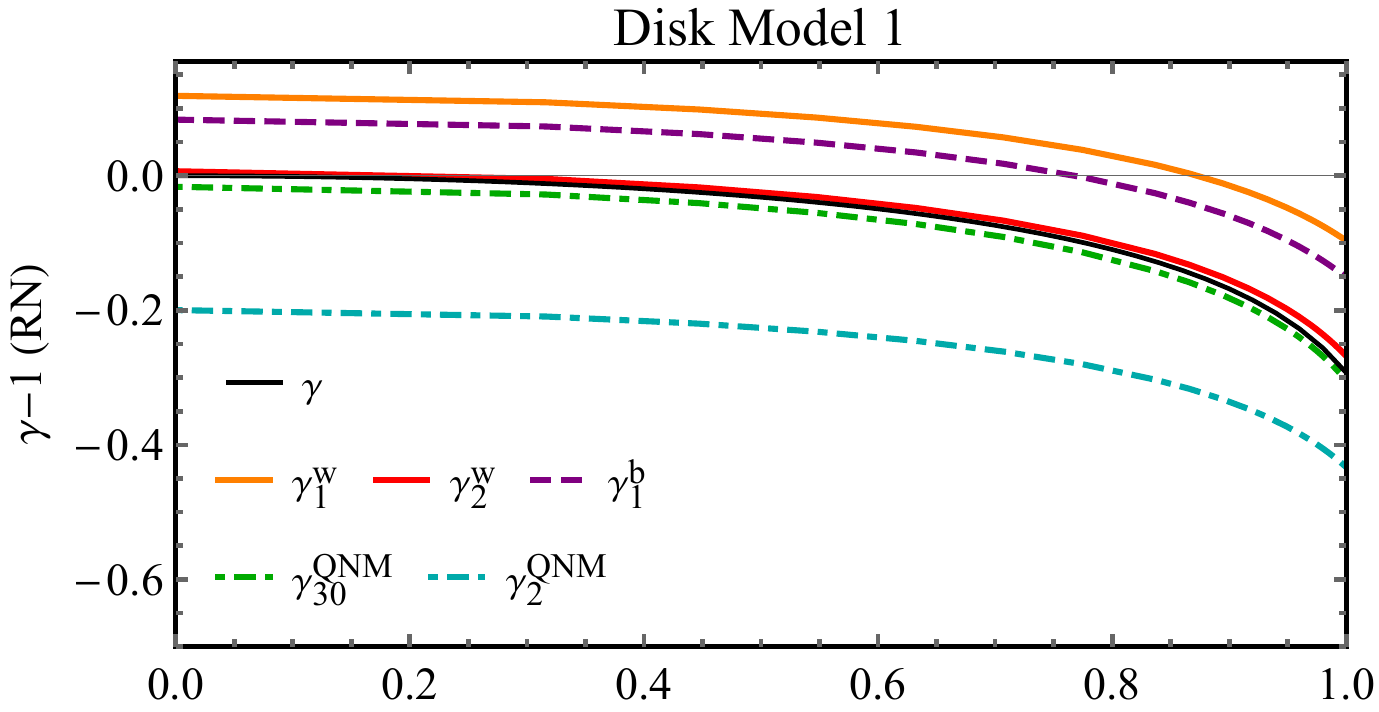}
    \includegraphics[scale=0.566]{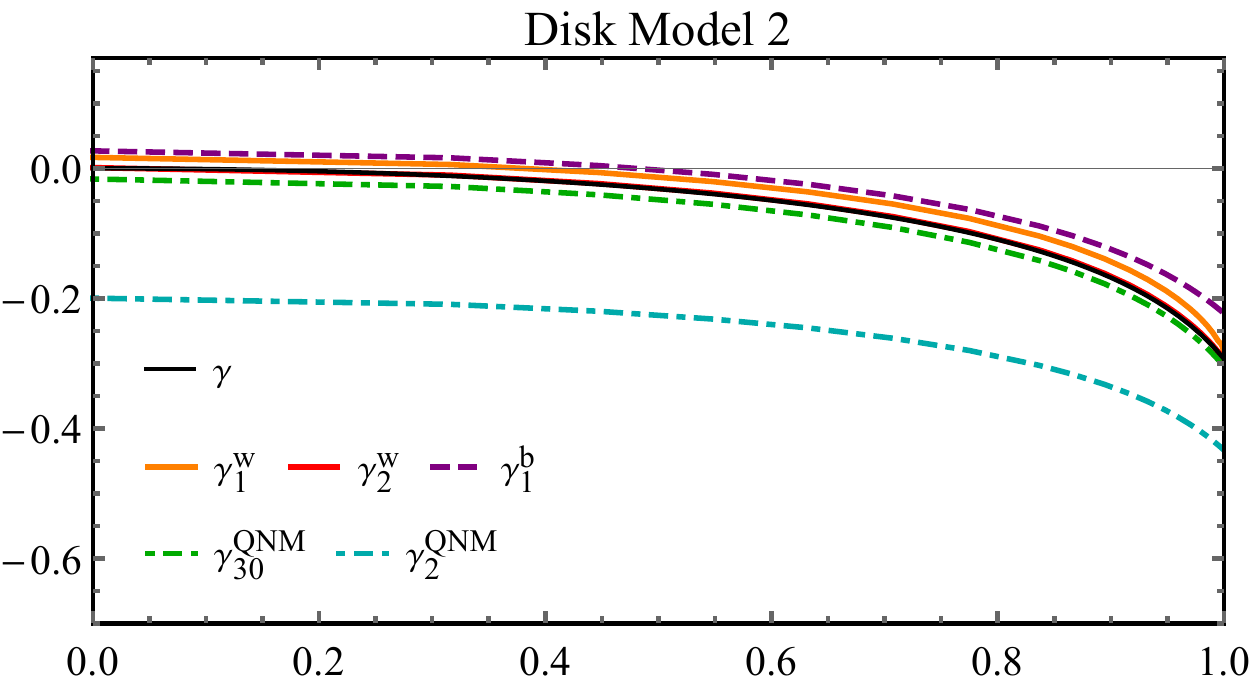}
    \includegraphics[scale=0.6]{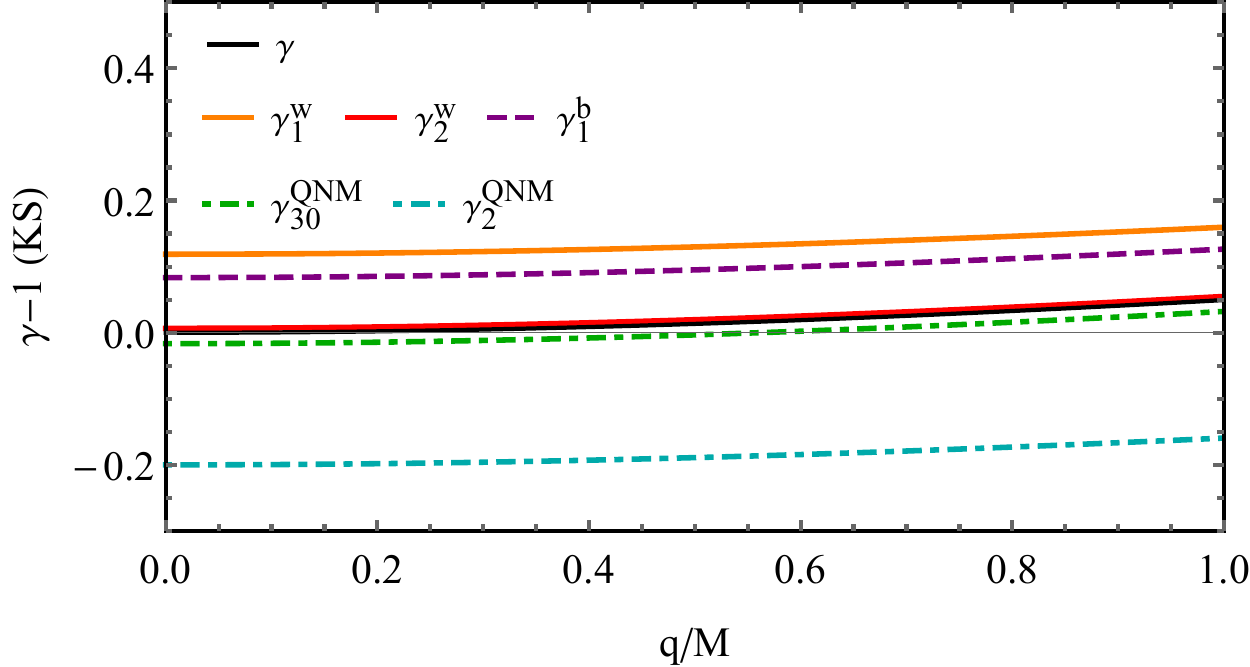}
    \includegraphics[scale=0.566]{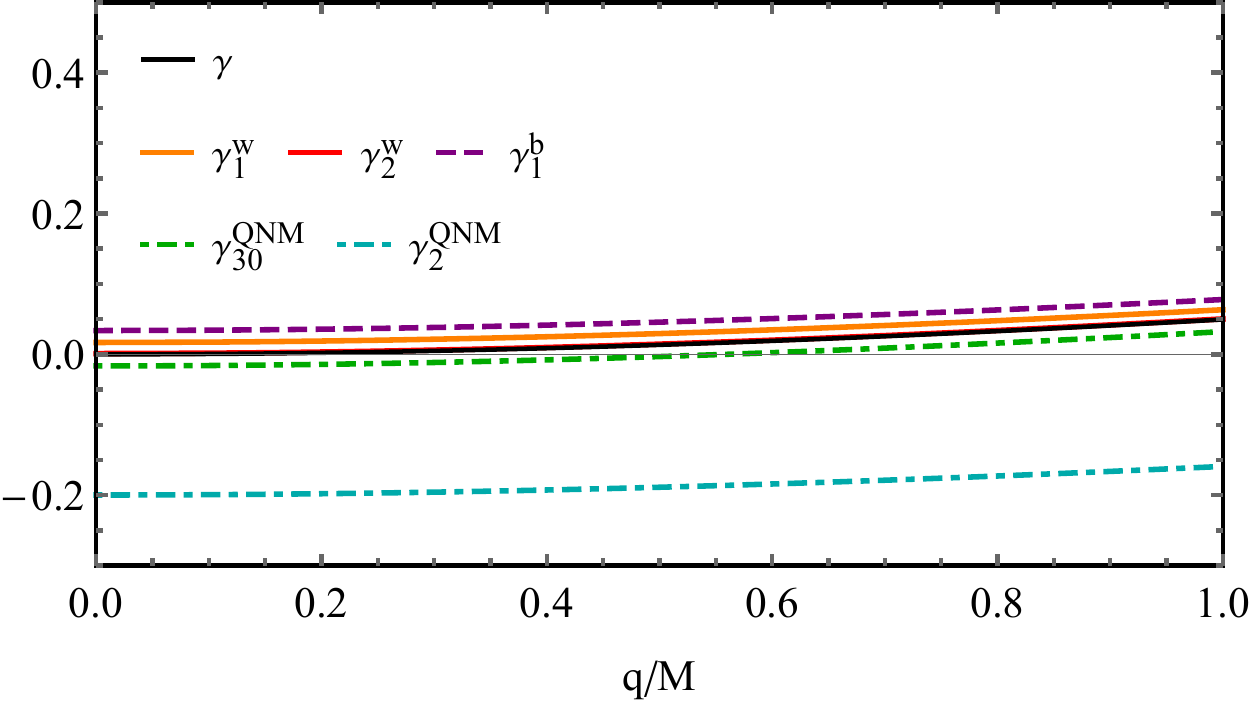}
    \caption{The critical exponent $\gamma$ (black), ring observables $\{\gamma_n^\textrm{w},\gamma_n^\textrm{b}\}$, and the QNM observables $\gamma_l^\textrm{QNM}$ for RN (upper) and KS (lower) spacetimes. Two disk models 1 (left) and 2 (right) are considered when evaluating the ring observables.}
    \label{fig:my_label}
\end{figure*}

We first consider the standard electrically charged and nonrotating black hole spacetime, i.e., the Reissner-Nordstr\"om (RN) spacetime, whose metric reads
\begin{equation}
f(r)=1-\frac{2M}{r}+\frac{q^2}{r^2}\,,\qquad g(r)=1\,,
\end{equation}
where $0\le q\le M$ is the charge. In the upper panels of Fig.~\ref{fig:my_label}, we consider disk models 1 (left) and 2 (right), and show the critical exponent $\gamma$ (black), the QNM observables of two different $l$ (dash-dotted), the ring observables $\gamma_n^\textrm{w}$ with $n=1$ (orange) and $n=2$ (red), and $\gamma_1^\textrm{b}$ (dashed), with respect to $q/M$ (we shift the vertical axis such that the black curve depicted by $\gamma$ vanishes in the Schwarzschild limit at $q=0$). One can see that, although the quantitative values of ring observables for a given $q$ depend on emission models, the observables, including the QNM ones, can all capture the qualitative behavior of $\gamma$ when changing $q$. In addition, in the presence of eikonal correspondence, the ring and QNM observables converge to $\gamma$ from opposite directions when $n$ and $l$ increase, with the former converging from above and the latter from below. Note that the observables $\gamma_2^\textrm{w}$ in both disk models are very close to $\gamma$. The red curves in the right panel are hardly seen because they are so close to the black ones.     

\subsection{Kazakov-Solodukhin black hole}

Then, we consider another non-Schwarzschild black hole spacetime and repeat the same calculations of observables as we just did for RN black holes. The black hole metric considered here is the Kazakov-Solodukhin (KS) one \cite{Kazakov:1993ha}
\begin{equation}
f(r)=-\frac{2M}{r}+\frac{\sqrt{r^2-q^2}}{r}\,,\qquad g(r)=1\,,
\end{equation}
where $q$ controls the deviations from the Schwarzschild metric. This spacetime can arise in a string-inspired model and can be regarded as a quantum-corrected Schwarzschild spacetime \cite{Kazakov:1993ha}. Roughly speaking, the quantum corrections in this model effectively shift the Schwarzschild singularity to a finite radius $q$. The QNMs \cite{Saleh:2014uca,Saleh:2016pke,Konoplya:2019xmn} and lensing effects \cite{Peng:2020wun,Lu:2021htd} of KS black holes have already been studied.

We present the results of KS black holes in the lower panels of Fig.~\ref{fig:my_label}. Although the critical exponent and observables increase with $q$ as opposed to the behavior we found in the RN case (upper panels), all the previous conclusions hold here. Namely, all the observables can capture the behavior of $\gamma$, and the QNM ones and ring ones converge to $\gamma$ from opposite directions. 

The two examples (RN and KS black holes) shown so far indicate that, if eikonal correspondence is satisfied, the ring observables should be larger than QNM ones and this result seems to be insensitive to the specific metrics and emission models under consideration. Therefore, by comparing these two sets of observables, the eikonal correspondence can be directly tested. An explicit example will be shown in the next subsection.

\subsection{Dynamical Chern-Simons gravity}
In the previous two subsections, we have shown that although the QNM observables $\gamma_l^\textrm{QNM}$ and photon ring observables $\{\gamma_n^\textrm{w},\gamma_n^\textrm{b}\}$ all converge to $\gamma$ in some limits, the former approaches $\gamma$ from below while the other from above. In this subsection, we would like to demonstrate how to test the eikonal correspondence by directly comparing these two sets of observables. For demonstration, we will take the dynamical Chern-Simons (dCS) theory of gravity as an example. It should be emphasized that the method should be applicable generically to test the eikonal correspondence of black holes.

The dCS gravity is a modified theory of gravity with a parity-violating quadratic curvature correction added to the Einstein-Hilbert action \cite{Alexander:2021ssr}. The theory continually receives great attention in recent years due to both its theoretical consistency of fundamental origin \cite{Alexander:2009tp}, and, of greater interest for the purpose of this work, its parity-violating features. In the dCS gravity, the Chern-Simons correction is introduced via a nonminimally coupled scalar field $\vartheta$ with a canonical kinetic term. The Schwarzschild metric is still the static and spherically symmetric vacuum solution to the theory. Its polar QNMs are the same as those in GR, while
the axial QNMs are coupled with the scalar degree of freedom contributed by $\vartheta$ \cite{Cardoso:2009pk}. The coupled equations are
\begin{equation}
\left(\frac{d^2}{dr_*^2}+\omega^2\right)
\begin{bmatrix}
    \Psi \\
    \Theta
    \end{bmatrix}=
    \begin{bmatrix}
    V_{11} & V_{12}\\
    V_{21} & V_{22}
    \end{bmatrix}
    \begin{bmatrix}
    \Psi \\
    \Theta
    \end{bmatrix}\,,\label{coupledeq1}
\end{equation}
where $\Psi$ stands for the axial metric perturbations, and $\Theta$ represents the perturbed scalar field. The components of the potential matrix are (we will set $\kappa=1/16\pi$)
\begin{align}
V_{11}&=f(r)\left[\frac{l(l+1)}{r^2}-\frac{6M}{r^3}\right]\,,\nonumber\\
V_{22}&=f(r)\left[\frac{l(l+1)}{r^2}\left(1+\frac{36M^2}{\kappa\beta r^6}\right)+\frac{2M}{r^3}\right]\,,\nonumber\\
V_{12}&=V_{21}=f(r)\sqrt{\frac{(l+2)!}{\beta\kappa(l-2)!}}\frac{6M}{r^5}\,,
\end{align}
where $f(r)=1-2M/r$ and $\beta$ is the dCS coupling constant. In this case, the QNM frequencies appear to have two branches labeled by $+$ and $-$, respectively. Up to the subleading eikonal order, the frequencies can be analytically expressed as \cite{Glampedakis:2019dqh} 
\begin{equation}
\omega_{\pm}=\omega_{R\pm}^{(0)}+\omega_{R\pm}^{(1)}+i\omega_{I\pm}+O\left(l^{-1}\right)\,
\end{equation}
with
\begin{equation}
    \begin{aligned}
    \omega_{R\pm}^{(0)}&=\left.\frac{l}{\sqrt{2}}\left[\tilde{V}(1+H)\pm\sqrt{\tilde{V}^2(1-H)^2+4\tilde{W}} \right]^{1/2}\right|_{r_m}\,,\\
    \omega_{R\pm}^{(1)}&=\left.
    \frac{l\tilde{V}\left[\omega^2(1+H)-2l^2\tilde{V}H\right]+W^{(1)}}
    {{2\omega(2\omega^2-l^2\tilde{V}(1+H))}}
        \right|_{r_m,\,\omega_{R\pm}^{(0)}}\,,\\
    \omega_{I\pm}&=\left.
    -\frac{f(r)}{2\sqrt{2}\omega l}
    \frac{\lvert V''_{\textrm{eff}}(r,\omega)\rvert^{1/2}}
    {\left[\tilde{V}^2(1-H)^2+4\tilde{W}\right]^{1/4}}
        \right|_{r_m,\,\omega_{R\pm}^{(0)}}\,,
    \end{aligned}
\end{equation}
where 
\begin{equation}
    \begin{aligned}
    H&=1+\frac{36M^2}{\kappa \beta r^6}\,,\quad
    \tilde{V}=\frac{f(r)}{r^2}\,,\\
    V_{\textrm{eff}}(r,\omega)&=l^2\tilde{V}\left[\omega^2(1+H)-l^2\tilde{V}H\right]+l^4\tilde{W}\,.
    \end{aligned}
\end{equation}
On the above expressions, we have defined $W\equiv V_{12}V_{21}=l^4\tilde{W}+W^{(1)}+O\left(l^{2}\right)$, and $r_m$ for each branch satisfies $V_{\textrm{eff}}'(r_m,\omega_{\pm})=0$.
Then, the QNM observables in this case can be defined as
\begin{equation}
    \gamma_l^{\pm}\equiv \frac{2l|\omega_{I\pm}|}{\omega_{R\pm}^{(0)}+\omega_{R\pm}^{(1)}}\,.
\end{equation}

\begin{figure}
    \centering  \includegraphics[scale=0.62]{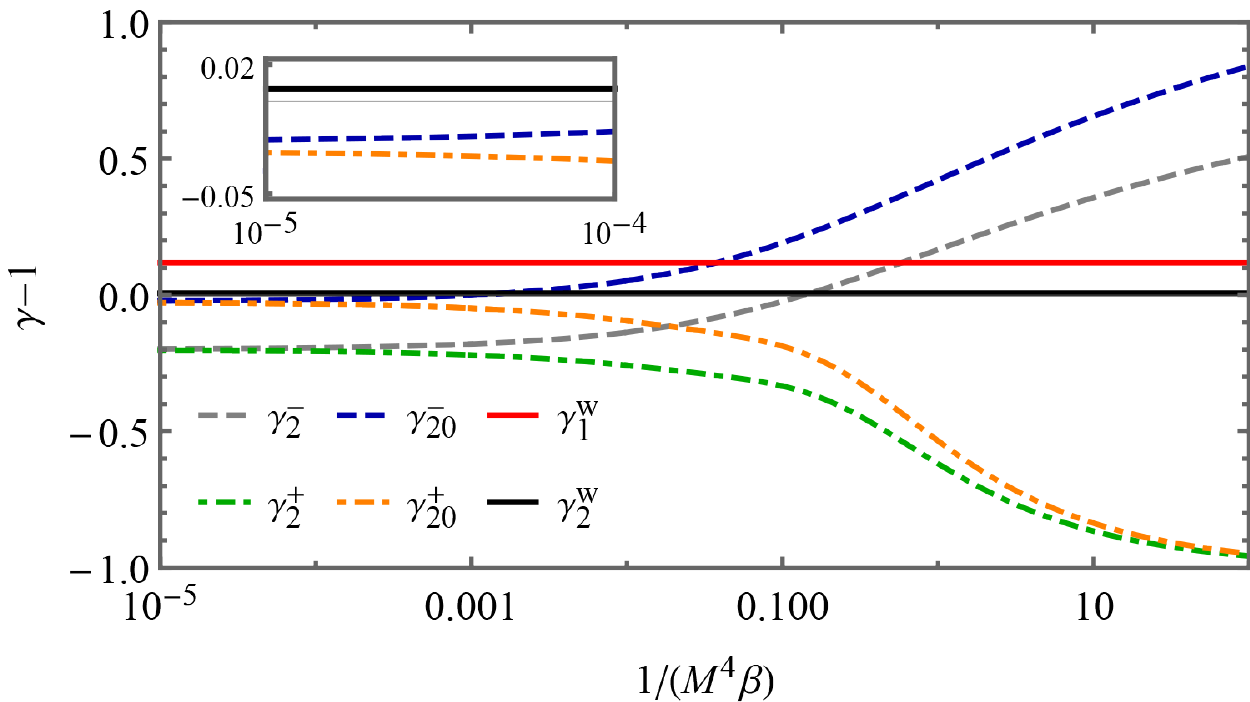}
    \caption{We show the two branches of QNM observables $\gamma_l^\pm$ for the Schwarzschild black hole in the dCS gravity as a function of $1/(M^4\beta)$ with $\ell=2$ (gray dashed and green dash-dotted) and $\ell=20$ (blue dashed and orange dash-dotted). The ring observables $\gamma_1^\textrm{w}$ (red) and $\gamma_2^\textrm{w}$ (black) with disk model 1 are shown for comparison. When $\beta$ decreases, the two branches separate and deviate from their GR values. In particular, $\gamma_l^-$ increases and may become larger than the ring observables at some $\beta$. Any indication of $\gamma_l^\pm$ being larger than ring observables is a clear signature of the violation of eikonal correspondence. The inset shows the details in the GR limits $M^4\beta\gg1$, in which the thin black line stands for $\gamma=1$.}     \label{dCS_Lyapunov}
\end{figure}

We choose $l=2$ and $l=20$, then show the two branches of $\gamma_l^\pm$ with respect to $1/(M^4\beta)$ in Fig.~\ref{dCS_Lyapunov}. As $M^4\beta\gg1$, two branches converge and reduce to their GR values. When $\beta$ decreases, the two branches separate, with the minus branch $\gamma_l^-$ increasing (dashed) while $\gamma_l^+$ decreases (dash-dotted). In this figure, we also show the ring observables $\gamma_1^\textrm{w}$ (red) and $\gamma_2^\textrm{w}$ (black) assuming disk model 1. As opposed to $\gamma_l^\pm$, the ring observables are blind to the change of $\beta$ because the spacetime metric and the photon geodesics are precisely those of the Schwarzschild spacetime. Any observational indication that the QNM observables are larger than the ring ones is a clear signature of the violation of eikonal correspondence. If the QNM observables were still observed to be smaller than ring ones, we can place constraints on the parameters that control the violation of eikonal correspondence, e.g. the parameter $\beta$ in the dCS gravity. This test based on the comparison between the QNM observables and ring ones is expected to be valid generically to test the eikonal correspondence of black holes.

\section{Conclusions}\label{sec.conclusion}

The eikonal correspondence between waves and photons, both propagating around a black hole, allows one to build a mapping between black hole QNMs and the optical appearance of the black hole. More explicitly, the real part of eikonal black hole QNMs can be mapped to the size of the critical curve in the image, and the imaginary part can be mapped to the detailed photon ring structures. This correspondence could be violated when going beyond GR. In this work, assuming the possibility of detecting the ringdown signals and of partially resolving the photon rings of a black hole, we propose a method specifically to test the eikonal correspondence. 

The method comprises the comparison of two sets of observables, the QNM observables \eqref{QNMgamma} and the photon ring observables \eqref{ringwb}. As the winding number $n$ increases, both two ring observables rapidly converge from above to the critical exponent $\gamma$ (Eq.~\eqref{criticalgamma}), which by itself is not an observable while can be purely determined by the metric. Considering nonrotating black hole spacetimes for illustration, we have shown that the way that the ring observables converge to $\gamma$ is insensitive to the astrophysical configuration around the black hole. On the other hand, if the eikonal correspondence is satisfied, the QNM observables converge to $\gamma$ from below when the multipole number $l$ increases. 

As we have exemplified using the RN and KS metrics, if the eikonal correspondence is fulfilled, all observables clearly capture the behavior of $\gamma$ when the spacetime deviates from the Schwarzschild one. The fact that they converge to $\gamma$ from opposite directions validates the test of eikonal correspondence simply by comparing these two sets of observables. If the QNM observables were measured to be larger than any of the ring ones, it could indicate the violation of eikonal correspondence. From a conservative perspective, one can also place constraints on the violation of the correspondence even if no ring observable was measured to be smaller than QNM ones. Furthermore, we would like to emphasize that the test would work even if only low-$l$ modes are extracted from gravitational wave signals. As shown in Fig.~\ref{fig:my_label}, the low-$l$ QNM observables already behave in the same trend as the critical exponent $\gamma$ and follow Eq.~\eqref{gammaqnm211} very well. As one can see from Fig.~\ref{dCS_Lyapunov}, the validity of the test does not strictly require the extraction of eikonal modes. However, certainly, tighter constraints can be placed if higher-$l$ modes can be extracted.

In this work, we have only considered the cases in which the effective potential $U(r)$ has a single peak. Whether the test proposed here can be extended to the effective potentials with multiple peaks deserves further investigation. An explicit example in Ref.~\cite{Guo:2021enm} already shows that the presence of multiple peaks with an inner global maximum, an outer local maximum, and a local minimum in between, would generate additional sets of modes in the QNM spectrum. The QNM frequencies corresponding to the global maximum would have the same expressions as those in the single-peak case given in Eqs.~\eqref{eq2.8} and \eqref{eq2.9}. If eikonal correspondence holds, these modes would correspond to the inner photon sphere and its associated image features \cite{Gan:2021pwu,Gan:2021xdl}. Therefore, we expect that the eikonal test should be applicable for these modes. On the other hand, the local minimum and maximum would lead to a set of long-lived modes and sub-long-lived modes \cite{Guo:2021enm}, respectively, which would contribute to the echos in time domain signals \cite{Guo:2022umh}. The long-lived modes are not detectable in the eikonal limit because of the infinitely high potential barrier. However, the real and the imaginary parts of sub-long-lived modes would correspond to the angular velocity and the Lyapunov exponent on the outer photon sphere, respectively, although the imaginary part is suppressed logarithmically by $l$ \cite{Guo:2021enm}. The eikonal test may be applicable for these sub-long-lived modes, possibly with a refined definition of the QNM observables.

So far, as a preliminary development of the test, we have only considered the face-on image of a nonrotating black hole lit by a thin disk. It is necessary to extend the present work by considering arbitrary inclination of the observer and rotating black holes. In both scenarios, the photon rings are not circular anymore and a more refined definition of ring observables is needed, although we do not expect changing inclination is going to alter our general conclusion. However, it is known that the Lyapunov exponent of bound photon orbits is sensitive to the spin of black holes. The validity of our method proposed here for rotating cases will be addressed elsewhere.

\appendix
\section{Analytic treatments of photon ring observables}\label{appendix1}
In this appendix, we will investigate the photon ring observables $\gamma_n^\textrm{w}$ and $\gamma_n^\textrm{b}$ using analytic treatments. The main goal is to show that 1) these observables converge to $\gamma$ as $n$ increases, and 2) they converge to $\gamma$ from above at least when the spacetime metric is sufficiently close to the Schwarzschild one.

Without loss of generality, we consider a photon trajectory that connects a light source at $r_s$ and a detector at $r_d$. The turning point where the trajectory is radially closest to the black hole is denoted as $r_0$. The consideration of the case without turning points is straightforward and we do not report it here. The change of angular coordinate $\phi$ along the trajectory can be calculated from the integral
\begin{equation}
    \Delta\phi=\sum_{i=s,d}\int_{r_0}^{r_i}\sqrt{\frac{g}{fr^2}}\frac{bdr}{\sqrt{\frac{r^2}{f}-b^2}}\,,\label{phiint}
\end{equation}
where $b$ is the impact parameter of the trajectory, $f$ and $g$ are the metric functions defined in Eq.~\eqref{metricsss}, and the turning point $r_0$ satisfies $b^2-r_0^2/f_0=0$. The impact parameter of the photon sphere, i.e., the radius of the critical curve on the ring image, is defined as $b_c\equiv\sqrt{r_{\textrm{ph}}^2/f_\textrm{ph}}$.

To proceed with the integration \eqref{phiint}, we define a new radial variable $z$ as \cite{Jia:2020qzt}
\begin{equation}
    z=1-b_c\sqrt{\frac{f}{r^2}}\,,\label{zr}
\end{equation}
such that the integral \eqref{phiint} can be rewritten as
\begin{equation}
    \Delta\phi=\sum_{i=s,d}\int_a^{z_i}\frac{F(z)}{\sqrt{(2-a-z)(z-a)}}dz\,,\label{phiint2}
\end{equation}
where $a\equiv1-b_c/b$, i.e., the $z$ at the turning point, and
\begin{equation}
    F(z)\equiv\sqrt{\frac{g(r(z))}{f(r(z))r(z)^2}}\left(1-z\right)\frac{dr}{dz}\,.\label{defF}
\end{equation}

We then consider the strong deflection limit where the turning point is close to the photon sphere, i.e., $a\rightarrow0^+$. Expanding $F(z)$ at $z=0$, we get
\begin{equation}
    F(z)=\sum_{n=-1}^\infty F_nz^{n/2}\,,\label{expF}
\end{equation}
where $F_n$ are coefficients. The integral \eqref{phiint2} can then be expressed as
\begin{align}
    \Delta\phi=&-\sqrt{2}F_{-1}\ln{a}-\frac{\sqrt{2}}{8}\left(3F_{-1}+4F_1\right)a\ln{a}\nonumber\\&+\sum_{i=s,d}X(z_i)+O(a)\,,\label{phiint3}
\end{align}
where
\begin{equation}
    X(z_i)=\frac{F_{-1}}{\sqrt{2}}\ln{\left(4z_i\right)}+F_0\sqrt{2z_i}+\frac{F_{-1}+4F_1}{4\sqrt{2}}z_i+o\left(z_i\right)\,.
\end{equation}
The constant term $\sum X(z_i)$ depends on the positions of the light source and the detector, and, as we will show later, does not contribute to the estimation of $\gamma_n^\textrm{w}$ and $\gamma_n^\textrm{b}$. Therefore, we will only focus on the coefficients $F_{-1}$ and $F_1$ which appear in the first two terms on the right-hand side of Eq.~\eqref{phiint3}.

Expanding $z$ with respect to $r-r_\textrm{ph}$, we get
\begin{equation}
    z\approx\frac{b_c^2}{4r_{\textrm{ph}}^4}\left(2f_{\textrm{ph}}-r_\textrm{ph}^2f''_{\textrm{ph}}\right)\left(r-r_\textrm{ph}\right)^2\,.
\end{equation}
Combining this expansion with Eqs.~\eqref{defF}
and \eqref{expF}, we get $F_{-1}=1/\sqrt{2}\gamma$. The expression for $F_1$ depends on higher derivatives of metric functions. It is very lengthy and not very informative, so we do not report it here.

Bringing in the relation between $\Delta\phi$ and the winding number, $\Delta\phi=(n+1/2)\pi$, one gets the lensing formula:
\begin{equation}
    -\gamma\left(n+1/2\right)\pi=\ln{a_n}+\epsilon_n-\ln{Y\left(z_s,z_d\right)}\,,\label{lsformula}
\end{equation}
where we have defined
\begin{equation}
a_n\equiv 1-\frac{b_c}{b_n}\,,\quad
    \epsilon_n\equiv \frac{\sqrt{2}\gamma}{8}\left(3F_{-1}+4F_1\right)a_n\ln{a_n}\,,
\end{equation}
and $Y\left(z_s,z_d\right)\equiv\textrm{exp}[\gamma\sum_{i=s,d}X\left(z_i\right)]$. The lensing formula \eqref{lsformula} is for the general metric \eqref{metricsss}, and it recovers those in the literature up to $O(a_n^0)$ \cite{Bozza:2002zj,Bozza:2007gt,Bozza:2010xqn,Tsukamoto:2016jzh}. Here we further extend the formula up to $O(a_n\ln{a_n})$, which, to the best of our knowledge, has not been done before. Note that $a_n$ is a decreasing function of the winding number $n$, and it vanishes when $n\rightarrow\infty$. Using the approximation $a_n\approx b_n/b_c-1$, we have
\begin{equation}
    b_n=b_c\left[1+\frac{Y\left(z_s,z_d\right)e^{-\gamma\left(n+1/2\right)\pi}}{1+\epsilon_n}\right]\,.\label{eqa11}
\end{equation}
With Eq.~\eqref{eqa11}, we estimate $b_n-b_{n+1}$ and the width of the $n$-th ring as
\begin{align}
b_n-b_{n+1}&=b_cY\left(z_s,z_d\right)e^{-\gamma\left(n+1/2\right)\pi}\nonumber\\
&\qquad\qquad\quad\times\left(\frac{1}{1+\epsilon_n}-\frac{e^{-\gamma\pi}}{1+\epsilon_{n+1}}\right)\,,\\
w_n&=b_c\left[Y(z_{so},z_d)-Y(z_{si},z_d)\right]\frac{e^{-\gamma\left(n+1/2\right)\pi}}{1+\epsilon_n}\,,
\end{align}
where $z_{so}$ and $z_{si}$ indicate that the light sources are located at the outer and the inner boundary of the disk, respectively. Then, from the above expressions, we can estimate the ring observable $\gamma_n^\textrm{b}$ of Eq.~\eqref{ringwb} as follows
\begin{align}
\gamma_n^\textrm{b}&\equiv\frac{1}{\pi}\ln{\frac{b_{n}-b_{n+1}}{b_{n+1}-b_{n+2}}}\nonumber\\&\approx\gamma+\frac{1}{\pi}\ln{\left[\frac{1-\epsilon_n-e^{-\gamma\pi}(1-\epsilon_{n+1})}{1-\epsilon_{n+1}-e^{-\gamma\pi}(1-\epsilon_{n+2})}\right]}\nonumber\\
&=\gamma+\frac{1}{\pi}\ln{\left[(1-e^{-\gamma\pi})\left(1+\frac{e^{-\gamma\pi}\epsilon_{n+1}-\epsilon_n}{1-e^{-\gamma\pi}}\right)\right]}\nonumber\\
&\qquad-\frac{1}{\pi}\ln{\left[(1-e^{-\gamma\pi})\left(1+\frac{e^{-\gamma\pi}\epsilon_{n+2}-\epsilon_{n+1}}{1-e^{-\gamma\pi}}\right)\right]}\nonumber\\
&\approx \gamma+\frac{\epsilon_{n+1}-\epsilon_n-e^{-\gamma\pi}\left(\epsilon_{n+2}-\epsilon_{n+1}\right)}{\pi\left(1-e^{-\gamma\pi}\right)}\,.
\end{align}
In addition, the observable $\gamma_n^\textrm{w}$ can be estimated as
\begin{align}
\gamma_n^\textrm{w}&\equiv\frac{1}{\pi}\ln{\frac{w_n}{w_{n+1}}}\nonumber\\&=\gamma+\frac{1}{\pi}\ln{\left(\frac{1+\epsilon_{n+1}}{1+\epsilon_{n}}\right)}\approx\gamma+\frac{\epsilon_{n+1}-\epsilon_n}{\pi}\,.
\end{align}
Recall that $a_n\ln{a_n}\rightarrow 0^{-}$ when $n\rightarrow\infty$. Therefore, the two observables both converge to $\gamma$ as $n$ increases. In addition, if the inequality $3F_{-1}+4F_1>0$ holds, such that $\epsilon_n\rightarrow0^{-}$ when $n\rightarrow\infty$, the observables converge to $\gamma$ from above. This inequality can be proven to be true for the Schwarzschild metric, and even for any nonrotating black hole spacetime that differs slightly from the Schwarzschild one. More explicitly, assume $g(r)=1+\delta g(r)$ and $f(r)=1-2M/r+\delta f(r)$. Up to $O(\delta f(r),\delta g(r))$ we have
\begin{align}
    &\frac{\sqrt{2}\gamma}{8}\left(3F_{-1}+4F_1\right)=\frac{5}{18}+\frac{1}{16}[16M\delta g'+60M^2\delta f''\nonumber\\&+12M^2\delta g''+96M^3\delta f'''+27M^4\delta f'''']\Big|_{r\rightarrow 3M}\,,
\end{align}
which is positive as long as the ``non-Schwarzschild" term, i.e., the second term on the right-hand side, is sufficiently small. In this case, the ring observables always converge to $\gamma$ from above, independent of emission models. In fact, our numerical calculations suggest that $3F_{-1}+4F_1$ is positive not only for RN and KS metrics when $q\approx 1$, but also for several other nonrotating black hole metrics, e.g. it is positive for all the asymptotically flat and nonrotating black holes considered in Ref.~\cite{Vagnozzi:2022moj}.

\acknowledgments
The authors thank Prashant Kocherlakota for his kind suggestions
on an earlier version of the manuscript. CYC is supported by the Institute of Physics of
Academia Sinica. MHO is
supported by the Ministry of Science and Technology, Taiwan under grant no. MOST 110-2112-M-001-068-MY3 and by Academia
Sinica, Taiwan under a career development award under grant no.
AS-CDA-111-M04. The initial stages of this work was done in part during the TCA Summer Student Program in 2021 and 2022 by the National Center for Theoretical Sciences.


\begin{thebibliography}{99} 

\bibitem{LIGOScientific:2016aoc}
B.~P.~Abbott \textit{et al.} [LIGO Scientific and Virgo],
Phys. Rev. Lett. \textbf{116}, no.6, 061102 (2016).

\bibitem{EventHorizonTelescope:2019dse}
K.~Akiyama \textit{et al.} [Event Horizon Telescope],
Astrophys. J. Lett. \textbf{875}, L1 (2019).

\bibitem{EventHorizonTelescope:2022wkp}
K.~Akiyama \textit{et al.} [Event Horizon Telescope],
Astrophys. J. Lett. \textbf{930}, no.2, L12 (2022).

\bibitem{Isi:2019aib}
M.~Isi, M.~Giesler, W.~M.~Farr, M.~A.~Scheel and S.~A.~Teukolsky,
Phys. Rev. Lett. \textbf{123}, no.11, 111102 (2019).

\bibitem{Johannsen:2010ru}
T.~Johannsen and D.~Psaltis,
Astrophys. J. \textbf{718}, 446-454 (2010).


\bibitem{Long:2020wqj}
F.~Long, S.~Chen, M.~Wang and J.~Jing,
Eur. Phys. J. C \textbf{80}, no.12, 1180 (2020).

\bibitem{Eichhorn:2021iwq}
A.~Eichhorn and A.~Held,
JCAP \textbf{05}, 073 (2021).

\bibitem{Delaporte:2022acp}
H.~Delaporte, A.~Eichhorn and A.~Held,
Class. Quant. Grav. \textbf{39}, no.13, 134002 (2022).

\bibitem{Chen:2020aix}
C.~Y.~Chen,
JCAP \textbf{05}, 040 (2020).

\bibitem{Fransen:2022jtw}
K.~Fransen and D.~R.~Mayerson,
Phys. Rev. D \textbf{106}, no.6, 064035 (2022).


\bibitem{Chen:2022lct}
C.~Y.~Chen,
Phys. Rev. D \textbf{106}, no.4, 044009 (2022).


\bibitem{Ferrari:1984zz}
V.~Ferrari and B.~Mashhoon,
Phys. Rev. D \textbf{30}, 295-304 (1984).

\bibitem{Hod:2009td}
S.~Hod,
Phys. Rev. D \textbf{80}, 064004 (2009).


\bibitem{Kokkotas:1999bd}
K.~D.~Kokkotas and B.~G.~Schmidt,
Living Rev. Rel. \textbf{2}, 2 (1999).

\bibitem{Berti:2009kk}
E.~Berti, V.~Cardoso and A.~O.~Starinets,
Class. Quant. Grav. \textbf{26}, 163001 (2009).

\bibitem{Konoplya:2011qq}
R.~A.~Konoplya and A.~Zhidenko,
Rev. Mod. Phys. \textbf{83}, 793-836 (2011).

\bibitem{Cardoso:2008bp}
V.~Cardoso, A.~S.~Miranda, E.~Berti, H.~Witek and V.~T.~Zanchin,
Phys. Rev. D \textbf{79}, no.6, 064016 (2009).

\bibitem{Dolan:2010wr}
S.~R.~Dolan,
Phys. Rev. D \textbf{82}, 104003 (2010).

\bibitem{Yang:2012he}
H.~Yang, D.~A.~Nichols, F.~Zhang, A.~Zimmerman, Z.~Zhang and Y.~Chen,
Phys. Rev. D \textbf{86}, 104006 (2012).

\bibitem{Li:2021zct}
P.~C.~Li, T.~C.~Lee, M.~Guo and B.~Chen,
Phys. Rev. D \textbf{104}, no.8, 084044 (2021).




\bibitem{Stefanov:2010xz}
I.~Z.~Stefanov, S.~S.~Yazadjiev and G.~G.~Gyulchev,
Phys. Rev. Lett. \textbf{104}, 251103 (2010).

\bibitem{Jusufi:2019ltj}
K.~Jusufi,
Phys. Rev. D \textbf{101}, no.8, 084055 (2020).

\bibitem{Jusufi:2020dhz}
K.~Jusufi,
Phys. Rev. D \textbf{101}, no.12, 124063 (2020).

\bibitem{Cuadros-Melgar:2020kqn}
B.~Cuadros-Melgar, R.~D.~B.~Fontana and J.~de Oliveira,
Phys. Lett. B \textbf{811}, 135966 (2020).

\bibitem{Yang:2021zqy}
H.~Yang,
Phys. Rev. D \textbf{103}, no.8, 084010 (2021).

\bibitem{Assumpcao:2018bka}
T.~Assumpcao, V.~Cardoso, A.~Ishibashi, M.~Richartz and M.~Zilhao,
Phys. Rev. D \textbf{98}, no.6, 064036 (2018).

\bibitem{Chen:2022ynz}
C.~Y.~Chen, H.~W.~Chiang and J.~S.~Tsao,
Phys. Rev. D \textbf{106}, no.4, 044068 (2022).




\bibitem{Chen:2021cts}
C.~Y.~Chen, M.~Bouhmadi-L\'opez and P.~Chen,
Eur. Phys. J. Plus \textbf{136}, no.2, 253 (2021).

\bibitem{Chen:2019dip}
C.~Y.~Chen and P.~Chen,
Phys. Rev. D \textbf{101}, no.6, 064021 (2020).

\bibitem{Nomura:2021efi}
K.~Nomura and D.~Yoshida,
Phys. Rev. D \textbf{105}, no.4, 044006 (2022)





\bibitem{Konoplya:2017wot}
R.~A.~Konoplya and Z.~Stuchl\'\i{}k,
Phys. Lett. B \textbf{771}, 597-602 (2017).

\bibitem{Glampedakis:2019dqh}
K.~Glampedakis and H.~O.~Silva,
Phys. Rev. D \textbf{100}, no.4, 044040 (2019).

\bibitem{Silva:2019scu}
H.~O.~Silva and K.~Glampedakis,
Phys. Rev. D \textbf{101}, no.4, 044051 (2020).

\bibitem{Moura:2021eln}
F.~Moura and J.~Rodrigues,
Phys. Lett. B \textbf{819}, 136407 (2021).

\bibitem{Bryant:2021xdh}
A.~Bryant, H.~O.~Silva, K.~Yagi and K.~Glampedakis,
Phys. Rev. D \textbf{104}, no.4, 044051 (2021).


\bibitem{Guo:2021enm}
G.~Guo, P.~Wang, H.~Wu and H.~Yang,
JHEP \textbf{06}, 060 (2022).





\bibitem{Li:2021mnx}
S.~Li, A.~A.~Abdujabbarov and W.~B.~Han,
Eur. Phys. J. C \textbf{81}, no.7, 649 (2021).




\bibitem{Gralla:2019xty}
S.~E.~Gralla, D.~E.~Holz and R.~M.~Wald,
Phys. Rev. D \textbf{100}, no.2, 024018 (2019).

\bibitem{Gralla:2019drh}
S.~E.~Gralla and A.~Lupsasca,
Phys. Rev. D \textbf{101}, no.4, 044031 (2020).




\bibitem{Blackburn:2019bly}
L.~Blackburn, S.~Doeleman, J.~Dexter, J.~L.~G\'omez, M.~D.~Johnson, D.~C.~Palumbo, J.~Weintroub, K.~L.~Bouman, A.~A.~Chael and J.~R.~Farah, \textit{et al.}
[arXiv:1909.01411 [astro-ph.IM]].









\bibitem{Johnson:2019ljv}
M.~D.~Johnson, A.~Lupsasca, A.~Strominger, G.~N.~Wong, S.~Hadar, D.~Kapec, R.~Narayan, A.~Chael, C.~F.~Gammie and P.~Galison, \textit{et al.}
Sci. Adv. \textbf{6}, no.12, eaaz1310 (2020).

\bibitem{Haworth:2019urs}
K.~Haworth, M.~D.~Johnson, D.~W.~Pesce, D.~C.~M.~Palumbo, L.~Blackburn, K.~Akiyama, D.~Boroson, K.~L.~Bouman, J.~R.~Farah and V.~L.~Fish, \textit{et al.}
[arXiv:1909.01405 [astro-ph.IM]].


\bibitem{Gralla:2020nwp}
S.~E.~Gralla,
Phys. Rev. D \textbf{102}, no.4, 044017 (2020).


\bibitem{Konoplya:2022gjp}
R.~A.~Konoplya,
[arXiv:2210.08373 [gr-qc]].





\bibitem{Wielgus:2021peu}
M.~Wielgus,
Phys. Rev. D \textbf{104}, no.12, 124058 (2021).

\bibitem{Broderick:2021ohx}
A.~E.~Broderick, P.~Tiede, D.~W.~Pesce and R.~Gold,
Astrophys. J. \textbf{927}, no.1, 6 (2022).

\bibitem{Kocherlakota:2022jnz}
P.~Kocherlakota and L.~Rezzolla,
Mon. Not. Roy. Astron. Soc. \textbf{513}, no.1, 1229-1243 (2022).

\bibitem{Kocherlakota:2022nhehttalk}
P.~Kocherlakota, (2022, June 22-25). Photon Rings in Spherically Symmetric Spacetimes [Conference presentation]. Assembling the ngEHT: Community-Driven Science to a Global Instrument, Granada, Spain.


\bibitem{Kocherlakota:2023qgo}
P.~Kocherlakota, L.~Rezzolla, R.~Roy and M.~Wielgus,
[arXiv:2307.16841 [gr-qc]].



\bibitem{Schutz:1985km}
B.~F.~Schutz and C.~M.~Will,
Astrophys. J. Lett. \textbf{291}, L33-L36 (1985).

\bibitem{Konoplya:2003ii}
R.~A.~Konoplya,
Phys. Rev. D \textbf{68}, 024018 (2003).

\bibitem{Konoplya:2019hlu}
R.~A.~Konoplya, A.~Zhidenko and A.~F.~Zinhailo,
Class. Quant. Grav. \textbf{36}, 155002 (2019).





\bibitem{Kazakov:1993ha}
D.~I.~Kazakov and S.~N.~Solodukhin,
Nucl. Phys. B \textbf{429}, 153-176 (1994).

\bibitem{Saleh:2014uca}
M.~Saleh, B.~B.~Thomas and T.~C.~Kofane,
Astrophys. Space Sci. \textbf{350}, no.2, 721-726 (2014).


\bibitem{Saleh:2016pke}
M.~Saleh, B.~T.~Bouetou and T.~C.~Kofane,
Astrophys. Space Sci. \textbf{361}, no.4, 137 (2016).

\bibitem{Konoplya:2019xmn}
R.~A.~Konoplya,
Phys. Lett. B \textbf{804}, 135363 (2020).

\bibitem{Peng:2020wun}
J.~Peng, M.~Guo and X.~H.~Feng,
Chin. Phys. C \textbf{45}, no.8, 085103 (2021).

\bibitem{Lu:2021htd}
X.~Lu and Y.~Xie,
Eur. Phys. J. C \textbf{81}, no.7, 627 (2021).




















\bibitem{Alexander:2021ssr}
S.~Alexander, G.~Gabadadze, L.~Jenks and N.~Yunes,
Phys. Rev. D \textbf{104}, no.6, 064033 (2021).

\bibitem{Alexander:2009tp}
S.~Alexander and N.~Yunes,
Phys. Rept. \textbf{480}, 1-55 (2009).

\bibitem{Cardoso:2009pk}
V.~Cardoso and L.~Gualtieri,
Phys. Rev. D \textbf{80}, 064008 (2009)




\bibitem{Gan:2021pwu}
Q.~Gan, P.~Wang, H.~Wu and H.~Yang,
Phys. Rev. D \textbf{104}, no.2, 024003 (2021).

\bibitem{Gan:2021xdl}
Q.~Gan, P.~Wang, H.~Wu and H.~Yang,
Phys. Rev. D \textbf{104}, no.4, 044049 (2021).

\bibitem{Guo:2022umh}
G.~Guo, P.~Wang, H.~Wu and H.~Yang,
JHEP \textbf{06}, 073 (2022).






\bibitem{Jia:2020qzt}
J.~Jia and K.~Huang,
Eur. Phys. J. C \textbf{81}, no.3, 242 (2021).









\bibitem{Bozza:2002zj}
V.~Bozza,
Phys. Rev. D \textbf{66}, 103001 (2002).


\bibitem{Bozza:2007gt}
V.~Bozza and G.~Scarpetta,
Phys. Rev. D \textbf{76}, 083008 (2007).

\bibitem{Bozza:2010xqn}
V.~Bozza,
Gen. Rel. Grav. \textbf{42}, 2269-2300 (2010).

\bibitem{Tsukamoto:2016jzh}
N.~Tsukamoto,
Phys. Rev. D \textbf{95}, no.6, 064035 (2017).




\bibitem{Vagnozzi:2022moj}
S.~Vagnozzi, R.~Roy, Y.~D.~Tsai, L.~Visinelli, M.~Afrin, A.~Allahyari, P.~Bambhaniya, D.~Dey, S.~G.~Ghosh and P.~S.~Joshi, \textit{et al.}
[arXiv:2205.07787 [gr-qc]].













\end{thebibliography}
\end{document}